# On melting of boron phosphide under pressure

Vladimir L. Solozhenko* and Vladimir A. Mukhanov

*LSPM–CNRS, Université Paris Nord, 93430 Villetaneuse, France*

E-mail: vladimir.solozhenko@univ-paris13.fr

Melting of cubic boron phosphide, BP has been studied at pressures to 9 GPa using synchrotron X-ray diffraction and electrical resistivity measurements. It has been found that above 2.6 GPa BP melts congruently, and the melting curve exhibits negative slope (-60±7 K/GPa), which is indicative of a higher density of the melt as compared to the solid phase.

Cubic boron phosphide, BP is refractory wide bandgap semiconductor [1,2] characterized by a unique combination of mechanical [3,4], thermal and electrical properties, excellent thermal conductivity [5] and high thermoelectric power [6] that could make it a material of choice for a wide range of engineering applications [7].

At ambient pressure decomposition of boron phosphide is observed already at 1400 K [8,9], and the melting point of BP is not established. The BP decomposition temperature is expected to be higher under pressure, however, so far the high-pressure behavior of boron phosphide is studied only at room temperature [10]. In the present work melting of boron phosphide at pressures to 9 GPa was studied for the first time.

Polycrystalline BP (99.8%) produced by self-propagating high-temperature reaction between boron phosphate and magnesium according to the method described elsewhere [4] was used in the experiments. The lattice parameter of the sample was $a = 4.5356(9)$ Å, which is close to the literature value (4.537 Å [9]).

The BP melting at ~5.3 GPa was studied *in situ* by energy-dispersive synchrotron X-ray diffraction using MAX80 multianvil system at F2.1 beamline of the DORIS III storage ring (HASYLAB-DESY). The experimental details are described elsewhere [11,12], the data obtained are shown by circles in Fig. 1. The angle-dispersive X-ray diffraction probing of BP melting at 6.2 and 8.9 GPa was performed in a Paris-Edinburgh press with T-cup module at ID30 beamline, ESRF. The experimental details are described elsewhere [12]; the corresponding data are presented in Fig. 1 as triangles.

Melting of boron phosphide in the 2.6–7.7 GPa pressure range was studied *in situ* by electrical resistivity measurements [13] in a specially designed high-temperature cell [14] of a toroid-type high-pressure apparatus. The cell was pressure-calibrated at room temperature using phase transitions in Bi (2.55 and 7.7 GPa), PbSe (4.2 GPa), and PbTe (5.2 GPa). The temperature calibration under pressure was made using well-established reference points: melting of Si, NaCl, CsCl, Pt, Rh, $Al_2O_3$, Mo and Ni–Mn–C ternary eutectic. No signs of chemical interaction between BP and graphite electrical inputs were observed over the whole pressure – temperature range under study. The experimental data are presented by squares in Fig. 1.

The BP melting curve (dashed line obtained by the least-squares method from the results of all experiments) exhibits negative slope –60(7) K/GPa, which points to the higher density of BP melt as compared to the solid phase in the pressure range under study. Extrapolation of the melting line to the low-pressure region allows us to estimate the melting point of BP at ambient pressure as 2840(40) K. The lattice parameters of the samples quenched from different pressures and temperatures are very close to the literature value, and no lines of other phases ($B_{12}P_2$, boron, phosphorus, etc.) are present in the diffraction patterns, which indicate the congruent type of BP melting at pressures above 2.6 GPa.

Synchrotron X-ray diffraction experiments were carried out during beam time allocated to Project DESY-D-I-20090172 EC at HASYLAB-DESY and Proposal HS-2532 at ESRF. The authors thank Drs. Christian Lathe and Yann Le Godec for assistance. This work was financially supported by the Agence Nationale de la Recherche (grant ANR-2011-BS08-018) and DARPA (Grant No. W31P4Q1210008).

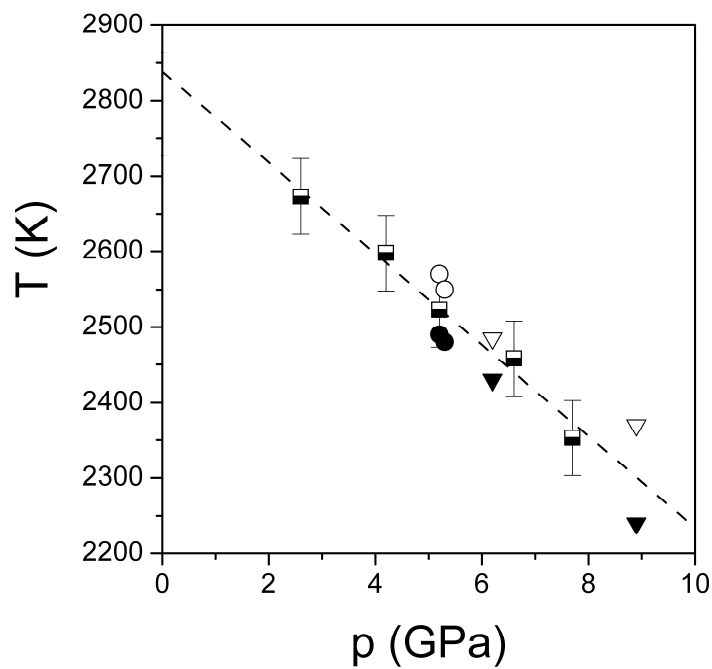

Fig. 1 Pressure dependence of BP melting temperature. The results of synchrotron X-ray diffraction experiments are presented by circles (HASYLAB-DESY) and triangles (ESRF). The open symbols correspond to the melt, solid symbols – to its absence. Half-filled squares indicate the onset of melting registered *in situ* by electrical resistivity measurements. Dashed line is the linear approximation of the melting curve defined by least-squares method.